\documentstyle[psfig]{aa}           % LaTeX A&A  Standard Fonts

\def\3he{$^3$He}
\def\12sur13{$^{12}$C/$^{13}$C}
\def\Msun{M$_{\odot}$}

\begin{document}
\thesaurus{ 
02.08.1; % Hydrodynamics
08.09.3, % Stars: interiors
08.18.1. % Stars: rotation
          }
\title{How many low-mass stars do destroy \3he?
}
\author{
C. Charbonnel
\,and
J.D. do Nascimento Jr 
}
\offprints{Corinne.Charbonnel@obs-mip.fr}
\institute{Laboratoire d'Astrophysique de Toulouse, UMR 5572 CNRS, 16 Avenue
E. Belin, 31400 Toulouse, France}

\date{Received March, 1998; accepted May 18, 1998}

\maketitle
\markboth{Charbonnel \& do Nascimento : $^3$He destruction in low-mass stars}{}

\begin{abstract}
{We recall the current status of the long-standing \3he problem, and its
possible connection with chemical anomalies on the red giant branch.
In this context, we collect in the literature all the available observations of 
the carbon isotopic ratio in field and cluster giant stars. 
Using the HIPPARCOS parallaxes, we get constraints on the 
evolutionary status of the field stars of the sample.
This allows us to identify the stars that have passed the 
luminosity function bump and present \12sur13 ratios in disagreement
with the standard predictions of stellar evolutionary models. 
We determine statistically what fraction of low mass stars 
experience an extra-mixing process on the red giant branch, and are then 
expected to destroy their \3he at this evolutionary phase. 
The high number we get satisfies the galactic requirements for 
the evolution of the \3he abundance.
}
\end{abstract}

\section{The $^3$He problem}
The evolution of \3he in the Galaxy has first been considered to be 
straightforward, 
dominated by the net production of this light element by low mass stars 
(i.e., with masses lower than 2 \Msun). In these objects, initial D is 
processed to \3he during the pre-main sequence phase. Then, as described
by Iben (1967), an \3he peak builds up due to pp-reactions on the main
sequence, and is engulfed in the stellar convective envelope during 
the first dredge-up on the lower red giant branch (RGB). 
Standard theory predicts that, once in the convective layers of 
the evolved star, \3he can not be destroyed because of the too cool 
temperature in these regions. 
It is finally ejected in the interstellar medium (ISM) in the latest 
stages of stellar evolution.
In this standard view, the abundance of \3he 
must increase in the Galaxy as soon as low mass stars begin to polute 
the ISM  (Rood et al. 1976).
One expects then to have constraints on the cosmological abundance of
\3he (Yang et al. 1984). 

Recent observations of a few planetary nebulae (PN; Rood et al. 1992,
Balser et al. 1997) led to the measurement of \3he in one object, 
NGC 3242
{\footnote {Balser et al. (1997) got no definitive detection of \3he
in any of the other five PN they observed; however detection of \3he 
is probable in two of their objects}}.
This PN, which estimated initial mass is 1.2$\pm$0.2 \Msun 
(Galli et al. 1997, hereafter GSTP97), presents a value of 
\3he/H$=(7.3 \pm 1.4) \times 10^{-4}$, in very good agreement with standard
predictions (Vassiliadis \& Wood 1993, Charbonnel 1995, Dearborn et al. 1996, 
Weiss et al. 1996 for the most recent computations).
This value however differs by almost two orders of magnitude with the 
\3he abundance in the proto-solar nebula, 
\3he/H=$(1.5 \pm 0.3) \times 10^{-5}$ (Geiss 1993), 
in the local interstellar cloud, 
\3he/H=$(2.2 \pm 0.2) \times 10^{-5}$ (Gloeckler \& Geiss 1996),
and in galactic HII regions, \3he/H=$(1~ {\rm to} ~5) \times 10^{-5}$
(Balser et al. 1994).

These low values are in clear contradiction with the conventional 
scenario for galactic evolution of the \3he abundance, and can not be explained
if all low mass stars, such as NGC 3242, happen to return all their \3he to 
the ISM.
GSTP97 showed that, in order to fit the galactic constraints, \3he should 
be destroyed in at least 70$\%$ of low-mass stars before they become PN.

\section{Constraints on the destruction of \3he in low-mass stars}
Rood et al. (1984) suggested that the destruction of \3he 
could be related to chemical anomalies such as the very low carbon isotopic 
ratios observed in low-mass red giants. 
Indeed, while standard models predict post-dredge up values of the \12sur13 
ratio between 20 and 30 (depending on the stellar mass and metallicity;
see Charbonnel 1994), 
Pop II field and globular cluster giants present \12sur13 ratios lower than 
10, even down to the near-equilibrium value of 4 in many cases.
This discrepancy also exists, but to a lower extent, in evolved stars 
belonging to open clusters with turnoff masses lower than 2 \Msun
(see Charbonnel et al. 1998, hereafter CBW98, for references). 

The \12sur13 data point out the existence of an extra-mixing process
that becomes efficient in low mass RGB stars as soon as they reach the 
so-called luminosity function bump (Gilroy \& Brown 1991, hereafter GB91; 
Charbonnel 1994; Pilachowski et al 1997; CBW98).
Different groups have simulated extra-mixing between the base of the 
convective envelope and the hydrogen-burning shell in order to reproduce
the chemical anomalies in RGB stars.  
These non-standard models predict that the mechanism 
which is responsible for the low \12sur13 ratios on the RGB must also lead 
to the destruction of \3he by a large factor in the bulk of the envelope 
material (Charbonnel 1995, Wasserburg et al. 1995, Weiss et al. 1996), and 
confirm the estimation of Hogan (1995).

In this paper, we use the observations of \12sur13 in evolved stars 
to determine statistically what fraction of low-mass stars experience 
this extra-mixing on the RGB, and are then expected to destroy their \3he.

\section{The data sample}
\subsection{Giants with measured \12sur13 ratios}

We collected in the literature all the available observations of \12sur13 ratios 
in stars of luminosity classes III and IV-III in the spectral ranges F to K.
The whole sample contains 191 stars, with [Fe/H] values between -2.8 and 0.3.  
Our 85 field stars have been observed by Lambert \& Ries (1981), Cottrell \& 
Sneden (1986), Sneden et al. (1986), Shetrone et al. (1993), Shetrone (1996), 
CBW98.
The data for galactic cluster giants are from GB91 for 
NGC 2682 (M67) and Gilroy (1989) for NGC 752, NGC 2360, IC 4756. 
For globular clusters, we use the data from Brown \& Wallerstein (1989) 
for NGC 104,  NGC 5139, NGC 6121 (M4), NGC 6656 (M22), 
from Smith \& Suntzeff (1989) for NGC 6121 and NGC 6656, 
from Suntzeff \& Smith (1991) for NGC 6121 and NGC 6752,
from Shetrone (1996) for NGC 6205 (M13), 
from Briley et al. (1997) for NGC 6838,
and from CBW98 for 47 Tuc.
When different values of the \12sur13 ratio are given by different authors 
for the same star, we take the most recent determination. 
We use the [Fe/H] values quoted in the respective papers. 

\subsection{Evolutionary status of our sample stars}

The determination of the \12sur13 ratio in our sample is not homogeneous.
This is however not a problem for our statistical study : 
We consider that a star has undergone extra-mixing when
its \12sur13 is smaller than 15, i.e., well below the minimum standard 
post dredge-up prediction. 

It is crucial however to have secure constraints on the evolutionary state 
of our sample stars. 
Indeed, as shown by GB91 and CBW98, the extra-mixing becomes efficient
on the RGB only after the luminosity function bump. 
Our statistics must thus be done only for stars that have already passed this
evolutionary state. 
The luminosity of the bump slightly depends on the initial metallicity.
In the [Fe/H] range we consider, M$_{\rm V}^{\rm bump}$
is higher than $\sim$ -0.2 (Fusi Pecci et al. 1990). 

For the field stars, we use the m$_{\rm V}$ and the parallaxes given 
in the HIPPARCOS catalogue to get M$_{\rm V}$.
For the globular cluster stars, we take the M$_{\rm V}$ values given 
in the respective papers quoted above. 
Precise absolute magnitudes depend on the cluster distance 
modulus, which is probably still uncertain. However, it matters only here 
to be sure that the cluster stars that we take into account in our final 
statistics are more luminous than the bump.

In our sample, we may have clump or AGB stars. Among the 106 cluster
giants, 10 objects have probably already reached this phase. However, 
their \12sur13 ratios does not differ (except in one case discussed in
\S 4.3) from the ones of the stars that are ascending the RGB for the 
first time and have passed the bump. 
Including them in the sample does not modify the final statistics. 

\begin{figure}
\psfig{figure=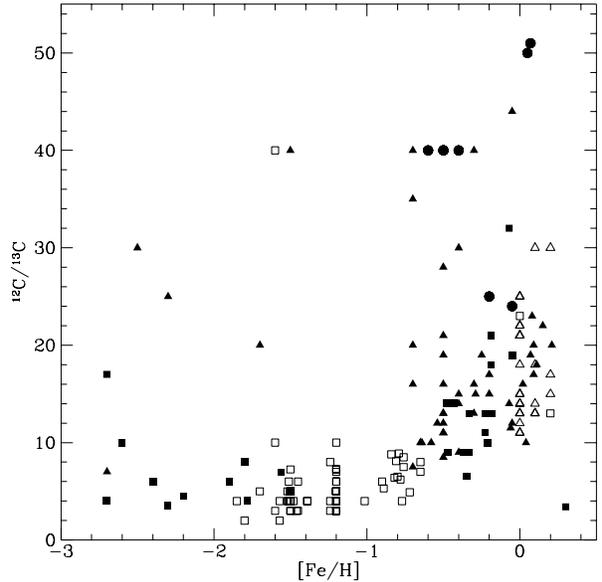,height=8.5cm,width=8.5cm}
\caption{Carbon isotopic ratio as a function of [Fe/H] for our whole sample. 
The circles correspond to stars with M$_{\rm V}$ higher than 2, the triangles 
to stars with M$_{\rm V}$ between 0 and 2, and the squares to the brightest 
stars with M$_{\rm V}$ lower than 0. Black and white symbols relate 
respectively to field and cluster giants}
\end{figure}

\section{Results}
\subsection{How many low-mass stars do destroy \3he?}
\begin{table}
\caption{Repartition of our sample stars in the three luminosity ranges
discussed in the text, as a function of their observed \12sur13 ratio}
\begin{tabular}{cccc}
\hline
 & \12sur13$\leq$15 & \12sur13$>$15 & Total \\
\hline
2$\leq$M$_{\rm V}$ & 0 & 8 & 8 \\
0$\leq$M$_{\rm V}<2$ & 35 & 40 & 75 \\
M$_{\rm V}<0$ & 101 & 7 & 108 \\
\hline
\end{tabular}
\end{table}

We show in Figure 1 the \12sur13 ratio as a function of [Fe/H] for all the
stars of our sample. The evolutionary status is also indicated by the 
M$_{\rm V}$ values. 
\begin{itemize}
\item Stars with M$_{\rm V}\geq2$ have not reached the bump yet, and they
all present \12sur13 ratios in agreement with standard predictions for 
dilution. They are excluded from our final statistics, since they are not
evolved enough to have undergone the RGB extra-mixing.  
\item For stars located in the bump region, i.e., with 0$\leq$M$_{\rm V}<$2, 
a large dispersion exists for the carbon isotopic ratio, which already 
appears to be very low in many giants. 
We do not take either these objects into account, since they are just in the 
region where they may experience the extra-mixing. 
\item Stars with M$_{\rm V} <$0 have passed the bump, i.e., the 
evolutionary point where the extra-mixing can occur. The disagreement 
between the standard predictions and the observations appears now in most
of the stars. 
\end{itemize}

The distribution of our sample stars in these three 
luminosity ranges is indicated in Table 1. 
In each domain, we give the number of stars which present ``normal" 
(i.e., higher than 15) and ``low"
(i.e., smaller than 15) carbon isotopic ratio. 

\begin{figure}
\psfig{figure=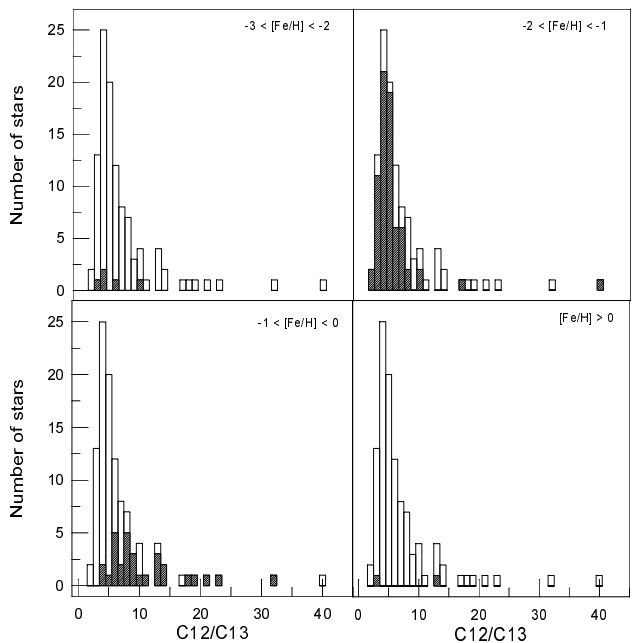,height=8.5cm,width=8.5cm}
\caption{Frequency distribution for the \12sur13 ratio for 
the stars that have already passed the luminosity bump, i.e., 
with M$_{\rm V}$ lower than 0. The shaded areas correspond 
to the different metallicity ranges indicated}
\end{figure}

The statistics we are interested in concern the most luminous stars. 
They are presented in the histograms of the Figure 2 as a function of 
their metallicity. 
We obtain that {\bf 93$\%$ of evolved stars undergo the extra-mixing on 
the RGB, and are thus expected to destroy, at least partly, their \3he}. 
This high number satisfies the galactic requirements, as discussed by 
GSTP97.
Let us note that if we take -0.5 as a limit for M$_{\rm V}$, we obtain
that {\bf 96$\%$ of evolved stars show a \12sur13 ratio in disagreement with
the standard predictions}.

In Figure 3, where we present the carbon isotopic ratios as a function of
M$_{\rm V}$ only for field stars, we underline the binary population
among our sample. Clearly, binaries and single stars have the same
behavior. 

\begin{figure}
\psfig{figure=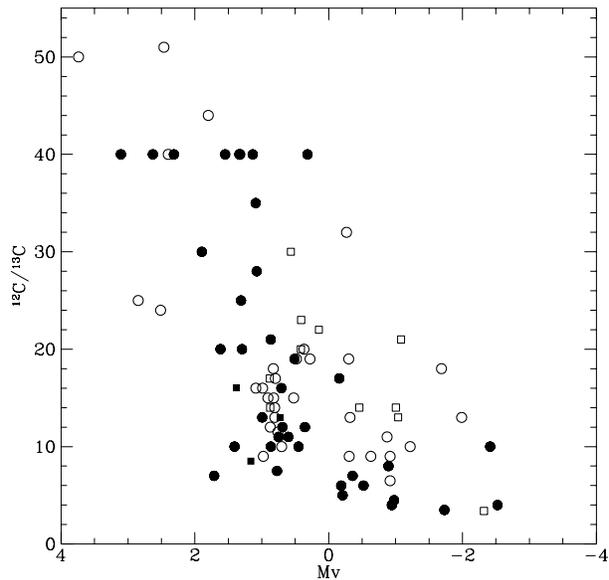,height=8.5cm,width=8.5cm}
\caption{Carbon isotopic ratio as a function of Mv for field stars.
The circles relate to single stars and the squares to binaries. 
Black and white symbols correspond respectively to [Fe/H] $\leq$-0.5 and 
[Fe/H]$>$ -0.5} 
\end{figure}

\subsection{Need for consistent yields}

The present non-standard models that attempt to explain the low \12sur13 
ratios all predict a severe destruction of \3he on the RGB. 
However, the task for stellar evolution theory is now to propose a 
physical process that explains consistently the various chemical anomalies 
observed in low mass red giant stars. 
Indeed, in addition to the \12sur13 problem, the behaviour of C, O, N, Al 
and Na on the RGB remains unexplained in many cases (see CBW98 for references). 
On the other hand, the extra-mixing process on the RGB has to destroy 
\3he in more than 90$\%$ of the low-mass stars, and preserve it in the others,  
for the high (and standard) \3he abundance observed in NGC 3242 to be 
explained. 
Stars with different histories (different rotation, mass loss, ...) could 
suffer different mixing efficiency and thus display different chemical 
anomalies. 
It is only when all these constraints will be explained that \3he yields 
by low mass stars will be reliable. 

\subsection{The ``deviant" stars}
In this context, one has to raise the question of the statistical
significance of Balser's et al. sample. As discussed by the
authors, this source sample is indeed highly biased, due to selection
criteria. The PN showing high \3he abundance should belong to the 
$\sim$7$\%$ of low-mass stars which do not suffer from extra-mixing 
on the red giant branch.
They should also show ``normal" carbon isotopic ratios.

This crucial test has already been verified for one PN 
of Balser's et al. sample for which \3he detection is probable: 
NGC 6720 shows a $^{12}$C/$^{13}$C ratio of 23 (Bachiller et al. 1997), 
in agreement with the ``standard" predictions. One has however to 
be cautious with this star for which the estimated initial mass is 
2.2$\pm$0.6 \Msun (GSTP97). This PN could have indeed been too 
massive to have reached the bump before igniting helium and thus to have
experienced the extra-mixing on the RGB. This is not the case for NGC
3242, which estimated initial mass is 1.2$\pm$0.2 \Msun. Observations of
$^{12}$C/$^{13}$C in this PN are now necessary.

Let us focus on the stars of the present sample that have passed the bump 
and do not behave as the majority. If we consider the objects with M$_{\rm V}$ 
lower than -0.5, only 3 stars appear to be really ``deviant" (see Table 2).

\begin{table}
\caption{Characteristics of the ``deviant" stars}
\begin{tabular}{ccccc}
\hline
Name & [Fe/H] & M$_{\rm V}$ & T$_{\rm eff}$ & \12sur13 \\
\hline
V8 (NGC 6656) & -1.6          & -5.62 & 4361 & 40 \\
HD 95689      & -0.19         & -1.08 & 4680 & 22 \\
HD 112989     & -0.44 to +0.3 & -2.32 & 4840 & 3.4 \\
\hline
\end{tabular}
\end{table}

\noindent {\bf V8 (NGC6656)}. ~
As can be seen in Fig.1 and 2, 
as soon as they are more luminous than the bump, almost all the stars
with [Fe/H] lower than -0.5 present carbon isotopic ratios lower than 10.
Only one star shows a \12sur13 ratio higher than the standard predictions : 
V8 in NGC 6656 (M22). This star presents enhanced SrII and BaII lines
(Mallia 1976) and is probably a star enriched in $^{12}$C and s-elements. 

\noindent {\bf HD 95689}($\alpha$Uma). ~
The relatively high Li abundance (logN(Li) = 1.26, Lambert et al. 1980) 
observed in this spectroscopic binary is another indication that this 
star did not suffer any extra-mixing on the RGB. It corresponds 
to the expected post-dilution Li for a star more massive than
1.7-2M$_{\odot}$ (see Charbonnel \& Vauclair 1992). 
This is consistent with the observed \12sur13 ratio and with the stellar
M$_{\rm bol}$ value we obtain with the HIPPARCOS data. For such an
initial stellar mass, the extra-mixing is not expected to occur on the RGB, 
and this star should be excluded from the present statistics. 

\noindent {\bf HD 112989}. ~
When [Fe/H] is higher than -0.5, the lower envelope of the \12sur13 ratio
lies around 12. 
One star however, HD 112989, lies well below this limit. 
For this binary star, important differences appear in the various 
estimations of [Fe/H] available in the literature (-0.44,
Yamashita 1964; +0.3, Lambert \& Ries 1981; +0.14, McWilliam 1990; 
-0.05, Dracke \& Lambert 1994). A smaller value of [Fe/H] would replace
this star in the ``normality". 

\begin{figure}
\psfig{figure=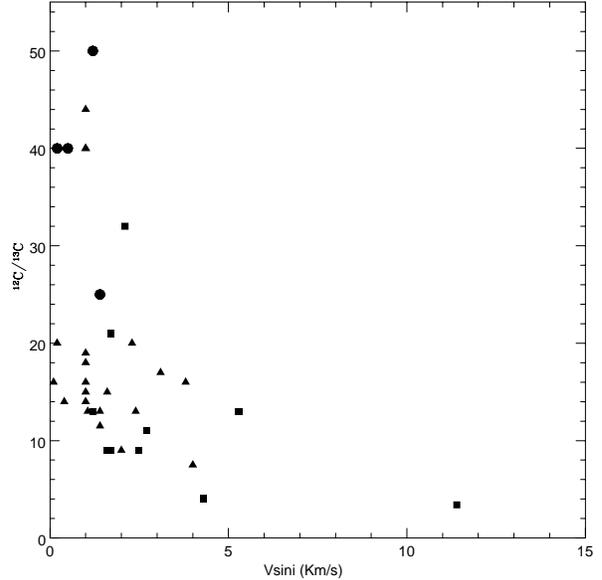,height=8.5cm,width=8.5cm}
\caption{Carbon isotopic ratio as a function of rotational velocity
for field stars.
The circles correspond to stars with M$_{\rm V}$ higher than 2, the
triangles to stars with M$_{\rm V}$ between 0 and 2, and the squares to
brightest stars with M$_{\rm V}$ lower than 0. 
HD112989 is the star with the highest Vsini}
\end{figure}

On the other hand, this weak-G band star presents a Vsini value of 
11km.sec$^{-1}$, which is much higher than the mean rotational velocity 
for giants with the same spectral type (De Medeiros 1990, 1998). 
Among the stars of our sample for which rotational velocity has been
measured with the CORAVEL spectrometer, this star is the fastest
rotator, all the others having Vsini lower than 5 km.sec$^{-1}$ (see
Fig.4).

\section{Conclusions}
We have assembled all the observations of the \12sur13 ratio in field
and cluster giants available in the literature. Using the HIPPARCOS
parallaxes, we get constraints on the evolutionary status of our sample
stars. We determine that 96$\%$ of low-mass stars do experience an
extra-mixing process on the RGB and are then expected to destroy their
\3he. While consistent ``non-standard" stellar models are needed to
explain the various chemical anomalies in low-mass RGB stars in order to
obtain reliable \3he yields, we can already conclude that the very high
percentage we get satisfies the galactic requirements for the evolution 
of the \3he abundance.

\begin{acknowledgements}
We thank J.R. de Medeiros for communicating us rotational velocity 
measurements obtained with the CORAVEL spectrometer.
We used in this work the SIMBAD data base and the Vizier tool
operated at the CDS (Strasbourg, France). J.D.N.Jr. acknowledges partial
financial support from the CNPq Brazilian Agency.
\end{acknowledgements}

\end{document}